\documentclass[10pt,conference]{IEEEtran}
\IEEEoverridecommandlockouts

\usepackage[most]{tcolorbox}
\usepackage{xspace}
\usepackage{enumitem}
\usepackage{pifont}
\usepackage{multirow,makecell}
\usepackage{color}
\usepackage[table, dvipsnames]{xcolor}
\usepackage{listings,amsfonts}
\usepackage{caption}
\usepackage{subcaption}
\usepackage{threeparttable}
\usepackage{bbding}
\usepackage{graphicx}
\usepackage{booktabs} 
\usepackage{longtable}
\usepackage{flushend}
\usepackage{tabularx}
\usepackage{siunitx} 
\usepackage{algorithm}
\usepackage{algpseudocode}
\usepackage{wasysym}
\usepackage{balance}

\newcommand{\full}{\CIRCLE}
\newcommand{\half}{\LEFTcircle}
\newcommand{\none}{\Circle}

\AtEndPreamble{
	\usepackage{hyperref}

}

\tcbuselibrary{listings,skins,breakable}

\definecolor{AFBoxBack}{HTML}{FBFBFB}
\definecolor{AFResolveBack}{HTML}{F7F9FC}
\definecolor{AFFactBack}{HTML}{FFFDF8}
\definecolor{AFFrame}{HTML}{C9C9C9}
\definecolor{AFTitleBack}{HTML}{F1F1F1}

\definecolor{AFCodeKey}{HTML}{520339}   % deeper purple
\definecolor{AFRefKey}{HTML}{10239e}    % deeper blue
\definecolor{AFFactKey}{HTML}{873800}   % deeper red-brown
\definecolor{AFText}{HTML}{202020}

\newcommand{\affont}{\ttfamily\fontsize{6.9pt}{7.85pt}\selectfont}
\newcommand{\afsectionfont}{\bfseries\fontsize{7.15pt}{7.9pt}\selectfont}

\lstdefinestyle{afsource}{
  basicstyle=\affont\color{AFText},
  columns=fullflexible,
  keepspaces=true,
  breaklines=true,
  breakatwhitespace=false,
  showstringspaces=false,
  tabsize=2,
  upquote=true,
  frame=none,
  numbers=none,
  aboveskip=0pt,
  belowskip=0pt,
  morekeywords={
    Agent,Runner,run,SQLiteSession,HostedMCPTool,
    function_tool,handoff,tools,handoffs,session,
    model,instructions,name,tool_config,server_label,server_url,
    require_approval,always,await,def,str,return
  },
  keywordstyle=\color{AFCodeKey},
  stringstyle=\color{AFText},
  commentstyle=\itshape\color{gray}
}

\lstdefinestyle{afalias}{
  basicstyle=\affont\color{AFText},
  columns=fullflexible,
  keepspaces=true,
  breaklines=true,
  breakatwhitespace=false,
  showstringspaces=false,
  tabsize=2,
  upquote=true,
  frame=none,
  numbers=none,
  aboveskip=0pt,
  belowskip=0pt,
  morekeywords={
    web_search,send_email,ops_agent,research_agent,session,
    model,instructions,tools,handoffs,entry,policy,state,
    capability,prompt,target,input
  },
  keywordstyle=\color{AFRefKey}
}

\tcbset{
  afboxbase/.style={
    enhanced,
    listing only,
    colframe=AFFrame,
    colbacktitle=AFTitleBack,
    coltitle=black,
    fonttitle=\bfseries\footnotesize,
    boxrule=0.35pt,
    arc=1pt,
    outer arc=1pt,
    left=1mm,
    right=1mm,
    top=0.75mm,
    bottom=0.75mm,
    boxsep=0.7mm,
    before skip=0pt,
    after skip=0pt
  }
}

\newtcblisting{sourcebox}[1][]{
  afboxbase,
  listing options={style=afsource,#1},
  title={1. Source Code},
  colback=AFBoxBack
}

\newtcblisting{resolvebox}[1][]{
  afboxbase,
  listing options={style=afalias,#1},
  title={2. Alias Resolution},
  colback=AFResolveBack
}

\newtcolorbox{factbox}{
  enhanced,
  title={3. Agent Facts},
  colback=AFFactBack,
  colframe=AFFrame,
  colbacktitle=AFTitleBack,
  coltitle=black,
  fonttitle=\bfseries\footnotesize,
  boxrule=0.35pt,
  arc=1pt,
  outer arc=1pt,
  left=1mm,
  right=1mm,
  top=0.75mm,
  bottom=0.75mm,
  boxsep=0.7mm,
  before skip=0pt,
  after skip=0pt
}

\newcommand{\fact}[1]{{\affont\textcolor{AFFactKey}{#1}}}
\newcommand{\obj}[1]{{\affont #1}}
\newcommand{\factsection}[1]{%
  \vspace{3pt}\noindent{\afsectionfont #1}\par\vspace{1pt}
}

\newcolumntype{F}[1]{>{\raggedright\arraybackslash\affont}p{#1}}

\newenvironment{factrecords}{
  \setlength{\tabcolsep}{0pt}
  \renewcommand{\arraystretch}{1.08}
  \begin{tabular}{@{}F{0.30\linewidth}F{0.36\linewidth}F{0.28\linewidth}@{}}
}{
  \end{tabular}
}

\newcommand{\toolname}{\textsc{AgentFlow}\xspace}

\begin{document}

\title{\toolname: Building Agent Dependency Graphs for Static Analysis of Agent Programs}

\author{
\IEEEauthorblockN{
Shenao Wang$^{\dagger}$,
Xinyi Hou$^{\dagger}$,
Yanjie Zhao$^{\dagger}$,
Xiao Cheng$^{\ddagger}$,
and Haoyu Wang$^{\dagger}$
}
\IEEEauthorblockA{
$^{\dagger}$Huazhong University of Science and Technology, \\
$^{\ddagger}$Macquarie University
}
\IEEEauthorblockA{
\{shenaowang, xinyihou, yanjie\_zhao, haoyuwang\}@hust.edu.cn,\\
jumormt@gmail.com
}
}

\maketitle

\begin{abstract}
LLM agents are increasingly developed as source-code applications built on agent frameworks. These \emph{agent programs} combine conventional host-language code with framework-defined semantics for models, prompts, tools, memory, and multi-agent orchestration logic. As a result, their behavior depends not only on traditional control and data flows, but also on a new class of \emph{agent dependencies}. Such dependencies are often expressed as framework-induced semantics, such as agent constructors, tool decorators, and agent handoff declarations, making them difficult to recover with existing static analysis or dependency tracking tools.
In this paper, we present \toolname, the first static analysis framework for recovering and analyzing agent dependencies from agent programs. \toolname constructs an \emph{Agent Dependency Graph}~(ADG), a framework-agnostic graph representation that represents agents, prompts, models, capabilities, memory states, and control policies as typed nodes, and captures their component-dependency, control-flow, and data-flow dependencies as typed edges. Built on ADGs, \toolname supports a range of analyses for agent governance and security, including Agent Bill of Materials~(BOM) generation and prompt-to-tool risk detection.
We implement \toolname for five representative agent frameworks and evaluate it on \textit{AgentZoo}, a corpus of 5,399 real-world agent programs. Our evaluation shows that \toolname recovers richer agent entities and dependencies than existing AST-based agent static analysis tools, generates more dependency-aware Agent BOMs, and uncovers 238 taint-style prompt-to-tool risks in real-world agent programs. These results show that ADG provides a practical foundation for understanding, governing, and securing emerging agent software.
\end{abstract}

\section{Introduction}

Large language models (LLMs) are increasingly integrated into autonomous agents that plan, invoke tools, access memory, and collaborate with other agents. This introduces a new programming paradigm in which software systems are no longer composed only of deterministic code logics, but also incorporate probabilistic LLM components that determine parts of the execution logic at runtime~\cite{owasp-agentic-ai-security-governance,jiang2026chaincaps,wang2026taintawi}. Modern agent frameworks such as LangGraph~\cite{langchain-langgraph}, OpenAI Agents~\cite{openai-agents-python}, and CrewAI~\cite{crewai} allow developers to build such applications by registering tools, specifying prompts, attaching memory, and orchestrating multi-agent workflows through framework-specific APIs. In this paper, we focus on source-code applications implemented through such frameworks, and refer to them as \emph{agent programs}\footnote{This scope excludes general agent products~(e.g., Claude Code, OpenClaw, and Hermes) and low-code agents~(e.g., GPT Store and Dify).}

Agent programs are framework-defined software built on ordinary host-language code. Beyond conventional program constructs such as functions, objects, classes, calls, assignments, and field accesses, agent frameworks introduce additional semantics through agents, tools, prompts, memory, and orchestration logic. These framework-induced semantics define a new class of agent-specific dependencies that classical dependency abstractions do not directly represent. We call them \emph{agent dependencies}. 
Traditional static analysis reasons about dependencies exposed by explicit program constructs, such as control flow~\cite{steven2014flowdroid,wei2014amandroid}, data flow~\cite{steven2014flowdroid}, call graphs~\cite{laursen2024approximate,wang2026yasa}, and points-to relations~\cite{sui2016svf,wang2026yasa}, where each dependency edge is derived at a syntactic site in the code. 
However, an agent dependency is often declared through framework-level constructs and spans heterogeneous agent-program entities~\cite{wang2026malskills,owasp-mas-threat-modeling-guide,pillar-not-all-aiboms}. For example, an agent constructor may appear as an ordinary host-language call, yet the framework interprets its prompt, tool, and memory arguments as agent-level dependencies, including data flow from prompts and memory to the agent and control/data flow between the agent and its invocable tools.
The question is therefore not merely whether one function may call another, but whether an agent may \emph{observe} certain information, \emph{propagate} it through prompts or memory, and \emph{use} it to trigger tools or other agents.

Modeling agent dependencies is necessary for several practical analyses of agent programs. First, it helps developers understand and threat-model agent programs by revealing which agents exist, which tools and resources they can access, how memory is shared, and where human or policy gates are enforced~\cite{owasp-agentic-ai-security-governance,melwin2026agentproof,agentwiz,owasp-mas-threat-modeling-guide}.
Second, it supports agent supply chain governance through Agent Bills of Materials~(BOMs). Traditional SBOM~\cite{spdx3,ntia-sbom-minimum-elements} and AI-BOM~\cite{karen2025aibom,vadim2025taibom,wiebe2026aibomgen} practices have become important mechanisms for improving supply chain transparency, compliance, and security, but they primarily inventory software packages, models, and datasets. Recent studies suggest that agent supply chains further introduce agent-native components, such as tools, MCP servers, and agent skills~\cite{pillar-not-all-aiboms,li2026agentbom,dutta2026agentriskbom}. These components may introduce tool poisoning~\cite{hou2025mcp,wang2026mcptox,hou2026mcpbiflow}, malicious skills~\cite{wang2026malskills,wu2026skillscope,guo2026malskillbench,liu2026skills}, and runtime composition risks~\cite{jiang2026chaincaps,wu2026chainfuzzer}. Therefore, Agent BOMs need to capture not only agent-specific components, but also their dependencies.
Third, it enables taint analysis for agent programs by tracking how untrusted inputs propagate through prompts, memory, and agents before reaching high-privilege tools. Prior work has studied prompt-to-tool injection vulnerabilities~\cite{liu2024llm4shell,liu2025agentfuzz,pedro2025prompt2sql,icse26taintp2x} and tool composition attacks~\cite{wu2026chainfuzzer,jiang2026chaincaps} in LLM-integrated applications. However, these works still follow traditional program analysis abstractions, focusing on code-level dependencies, rather than modeling agent-specific dependencies and taint propagations. Together, these needs motivate a systematic way to model agent dependencies in these emerging programs.

Unfortunately, resolving agent dependencies from source code is challenging. First, different agent frameworks expose heterogeneous programming models and different framework semantics. LangGraph represents agent behavior through stateful graph nodes and edges. OpenAI Agents exposes agents, tools, handoffs, guardrails, and sessions. CrewAI organizes agents, tasks, crews, tools, memory, and delegation policies. Second, agent dependencies are often implicit. For example, a tool may be registered by a decorator, a prompt may be assembled dynamically, a memory store may be hidden behind a framework abstraction, and a privileged action may be reachable only through an agent delegation chain. Third, the execution of an agent program is partly determined by the LLM components and their probabilistic outputs. Static analysis therefore cannot simply predict a single deterministic call sequence. Instead, it needs to approximate the execution trace permitted by the program architecture.

In this paper, we present \toolname, a static analysis framework for agent programs. The key idea of \toolname is to recover the \emph{Agent Dependency Graph} (ADG) as a unified Intermediate Representation~(IR) from heterogeneous agent framework implementations. An ADG models agents, prompts, models, external capabilities, memory states, and control policies as typed nodes. It captures their interactions as typed dependency edges, including component-dependencies, control-flow dependencies, and data-flow dependencies. Built on ADGs, \toolname supports a range of dependency-driven analyses. For governance, it constructs Agent BOMs that record not only agent components but also their semantic dependencies. For security, it supports taint-style prompt-to-tool risk analysis over the ADG, identifying agents that can reach privileged capabilities. These applications show that ADGs provide a common foundation for understanding, governing, and securing agent programs.

% We implement \toolname for XXX representative agent frameworks, including XXX. We evaluate it on XXX real-world agent programs and XXX benchmark applications. Our evaluation shows that \toolname constructs ADGs with XXX precision and XXX recall. It improves dependency extraction over existing inventory, workflow, and security scanning baselines by XXX\%. It also detects XXX previously unknown or seeded dependency risks. In addition, the generated ADGs support Agent BOM construction with XXX coverage of manually labeled agent components and dependencies.

In summary, we make the following contributions:

\begin{itemize}[leftmargin=15pt]

\item \textbf{Unified Agent IR.}
We propose the ADG as a framework-agnostic IR for agent programs, which represents semantic dependencies over agents, models, prompts, capabilities, memory state, and control policy in a unified graph.

\item \textbf{Practical Static Analyzer.}
We design and implement \toolname, the first static analyzer that recovers ADGs from agent program source code across heterogeneous framework constructs and abstractions.

\item \textbf{Applications and Findings.}
We demonstrate \toolname{} through two ADG-based applications, Agent BOM generation and taint-style prompt-to-tool risk detection on 5,399 real-world agent programs. The results show that \toolname{} generates more dependency-aware Agent BOMs than existing tools and uncovers 238 real-world prompt-to-tool risks.

\end{itemize}
\section{Background}
\label{sec:background}

\subsection{Agents and Agent Programs}
\label{sec:background-agents}

An agent is an LLM-driven software that uses an LLM to interpret context, make decisions, and perform actions through external capabilities. Unlike chat-style LLM applications that primarily produce text~\cite{yan2024gptapp,yan2025gptapp,hou2025llmapp,zhao2025llmapp}, an agent is typically embedded in an execution loop where model outputs can influence subsequent observations, tool invocations, state updates, and interactions with external environments. 
Recent protocols and specifications, such as MCP, A2A, and Agent Skills, further standardize how agents invoke tools, communicate with other agents, and reuse packaged capabilities~\cite{anthropic-mcp,google-a2a,anthropic-agent-skills}. As the agent paradigm is increasingly adopted, LLM agents have emerged in multiple application forms and implementation styles~\cite{owasp-agentic-ai-security-governance}. 
% As summarized in \autoref{tab:agent-forms}, agent applications can be roughly categorized into three forms. \emph{Agent products} are end-user-facing systems, such as coding agents, browser-use agents, and personal assistants. \emph{Agent platforms} allow users or developers to build agents through platform configurations or low-code workflows. In contrast, \emph{agent frameworks} provide SDKs and programming abstractions for implementing agent behavior directly in source code. 

% \begin{table}[t]
% \centering
% \caption{Representative agent application forms.}
% \label{tab:agent-forms}
% \begin{tabular}{cl}
% \toprule
% \textbf{Agent Type} & \textbf{Representative Forms} \\
% \midrule
% \multirow{4}{*}{\textbf{Products}}
% & Coding agents, e.g., Claude Code, Codex \\
% \cmidrule(lr){2-2}
% & Browser-use agents, e.g., OpenAI Operator \\
% \cmidrule(lr){2-2}
% & Personal assistants, e.g., OpenClaw, Hermes \\
% \midrule

% \multirow{2}{*}{\textbf{Platforms}}
% & Chat-style LLM apps, e.g., GPT Store, Coze \\
% \cmidrule(lr){2-2}
% & Low-code workflow agents, e.g., n8n, Dify \\
% \midrule

% \multirow{2}{*}{\textbf{Frameworks}}
% & Single-agent programs, e.g., LangChain \\
% \cmidrule(lr){2-2}
% & Multi-agent programs, e.g., CrewAI Crews \\
% \bottomrule
% \end{tabular}
% \end{table}

\noindent\textbf{Scope.} In this paper, we focus on agent programs built with agent frameworks. In this setting, source code is the primary analysis artifact, and provides the basis for recovering the agent dependencies. This distinguishes agent programs from agent products and low-code agents, such as coding agents, claw-like agents, or dify agents.

\subsection{Agent Frameworks and SDKs}
\label{sec:background-frameworks}

\begin{figure}[t]
\centering
\includegraphics[width=0.98\linewidth]{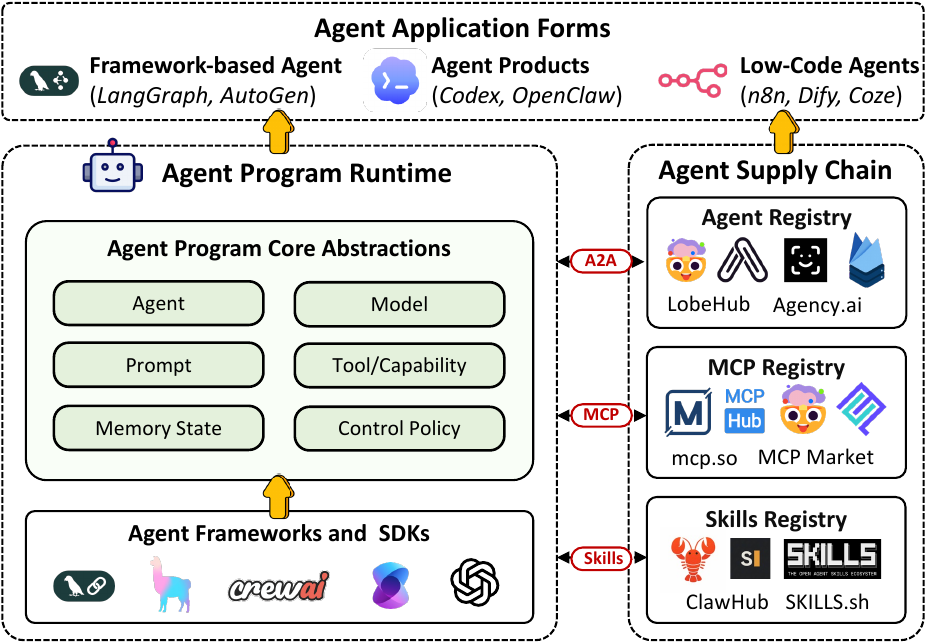}
\caption{Agent programs and supply chain components.}
\label{fig:agent-background}
\end{figure}

Agent programs are commonly implemented using agent frameworks and SDKs that provide programming abstractions for connecting LLMs with tools, memory, external resources, and multi-agent orchestration~\cite{xue2025agentframework,wang2025agentframework,ning2026defects}. Although these frameworks differ in their APIs and execution semantics, recent efforts such as Oracle's Agent Spec suggest that agent applications share a set of common concepts~\cite{amini2026agentspec}. As illustrated in \autoref{fig:agent-background}, these framework implementations can be abstracted into a common set of agent-program core entities, namely \emph{Agent}, \emph{Prompt}, \emph{Model}, \emph{Agent Capability}, \emph{Memory State}, and \emph{Control Policy}.

\begin{figure*}[t]
\centering
\includegraphics[width=0.98\linewidth]{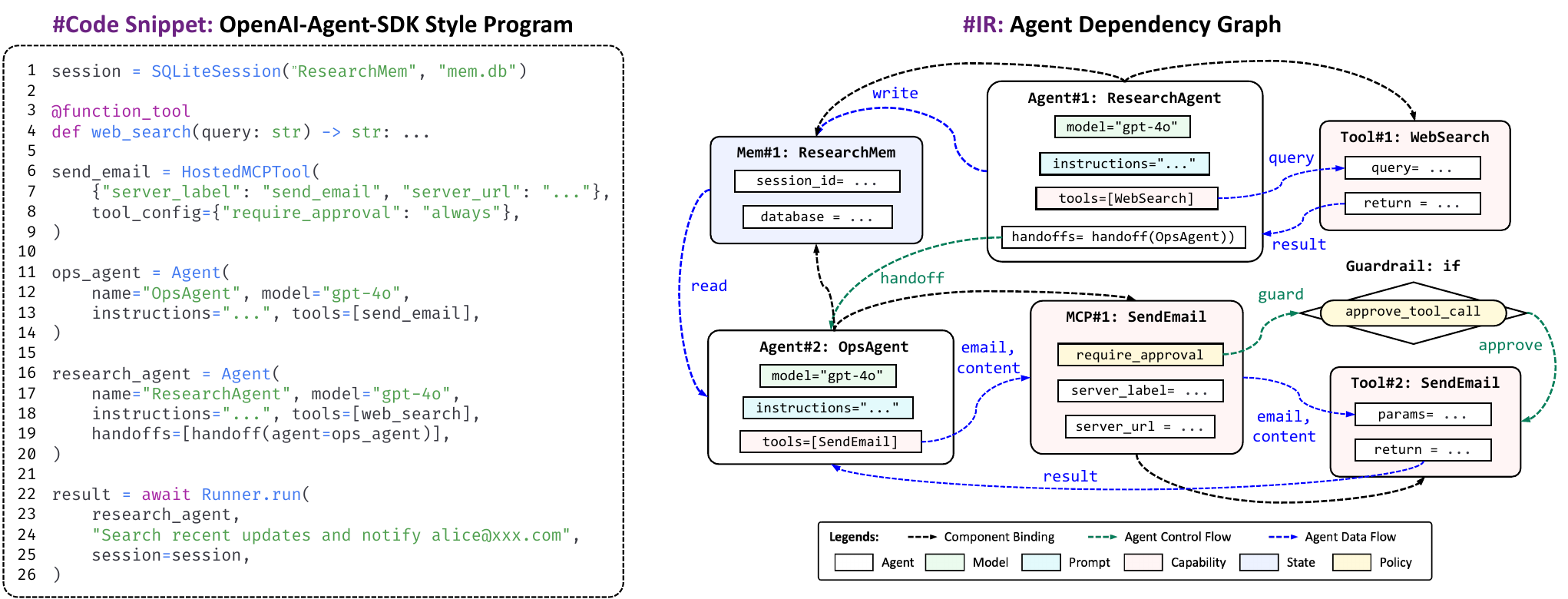}
\caption{Motivating example of agent-specific dependencies in an agent program. The left side shows an OpenAI-Agent-SDK-style program with a function tool, a hosted MCP tool, two agents, a shared session object, and an agent-to-agent handoff.}
\label{fig:motivating-agent-dependency-example}
\end{figure*}

\begin{itemize}[leftmargin=15pt]
    \item \textbf{Agent, Prompt, and Model.}
    An \emph{agent} is the primary execution unit. It is typically configured with a \emph{prompt} that specifies its role, task, and behavioral constraints, and a \emph{model} that determines which LLM is used for response generation. These abstractions define how the agent interprets context and produces outputs.

    \item \textbf{Agent Capability.}
    An \emph{agent capability} denotes an external action or reusable functionality that an agent can invoke during execution. It may take the form of built-in tools, user-defined tool functions, hosted MCP capabilities, or packaged skills. These capabilities define how an agent interacts with external systems and extend its operational boundary beyond model generation.

    \item \textbf{Memory State.}
    \emph{Memory state} allows agents to persist conversation history, retrieved knowledge, intermediate summaries, or long-term task context across turns. They are important for dependency analysis because information written in one step may later influence another agent or capability invocation.

    \item \textbf{Control Policy.}
    \emph{Control policies} constrain or mediate agent behavior, including human-in-the-loop approval, guardrails, permission checks, and validation constraints. They define the control points at which sensitive actions may be blocked, approved, or redirected.
\end{itemize}

\subsection{Agent Supply Chain and BOM}
\label{sec:background-supply-chain}

As agent programs become increasingly modular, their capabilities are extended by agent-specific supply chain components. Prior work on the LLM supply chain has shown that LLM-based systems are composed of heterogeneous artifacts, models, data, prompts, tools, plugins, and deployment environments~\cite{wang2025llmsc,hu2025llmsc,huang2025llmsc}. As shown in \autoref{fig:agent-background}, in agent programs, this composition further expands into an \emph{agent supply chain}, where third-party components, such as tools, MCP servers, and Agent Skills, can be published, discovered, installed, configured, and invoked by agents during execution~\cite{owasp-agentic-ai-security-governance,wang2026malskills}.
However, the agent supply chain also introduces new attack surfaces. Recent studies have shown that MCP tools can be abused through tool poisoning and unsafe tool descriptions~\cite{hou2025mcp,wang2026mcptox,hou2026mcpbiflow}, skills can contain malicious behavior distributed across prompts, code, manifests, and configurations~\cite{wang2026malskills,wu2026skillscope,guo2026malskillbench,liu2026skills}, and individually legitimate tools can be composed into unsafe tool chains~\cite{jiang2026chaincaps,wu2026chainfuzzer}. These risks suggest that the transparency needed for agent programs must go beyond identifying which components are present.

Existing SBOM practices inventory software packages, build artifacts, and vulnerability metadata, while AI-BOM practices extend inventory to models, datasets, environments, metadata, and provenance~\cite{spdx3,ntia-sbom-minimum-elements,karen2025aibom,vadim2025taibom,wiebe2026aibomgen}. An Agent BOM, however, should further capture agent-specific components and their dependency relations, such as which agents can invoke which tools, which tools access which resources, which skills or MCP servers extend agent capabilities, how memory propagates information, and which guards protect high-impact actions. Recent work on Agent BOM similarly suggests that security-auditable agent documentation requires visibility into capabilities, tool use, memory operations, and cross-agent propagation~\cite{li2026agentbom}. In this work, we support Agent BOM generation by statically analyzing agent program dependencies from source code.
\section{Motivating Example}
\label{sec:motivating-example}

\autoref{fig:motivating-agent-dependency-example} shows a simplified OpenAI-Agent-SDK-style program. The program first initializes a shared \texttt{SQLiteSession}~(L1), then defines a local function tool \texttt{web\_search}~(L3--L4) and a hosted MCP tool \texttt{send\_email}~(L6--L9). The MCP tool is configured with \texttt{require\_approval="always"}~(L8), indicating that the email-sending action is protected by an approval policy. It then defines two agents. The \texttt{OpsAgent}~(L11--L14) is backed by \texttt{gpt-4o} and equipped with \texttt{send\_email}. The \texttt{ResearchAgent}~(L16--L20) is equipped with \texttt{web\_search}, and configured with a handoff to \texttt{OpsAgent}~(L19). Finally, it runs \texttt{ResearchAgent} over a user request and attaches the session to share memories between the two agents~(L22--L25).

At the source-code level, the agent dependencies are introduced through framework-specific abstractions rather than ordinary procedure calls. In the example, \texttt{tools=[...]} exposes capabilities to agents, \texttt{handoff(...)} defines a possible transfer of execution from \texttt{ResearchAgent} to \texttt{OpsAgent}, \texttt{session=session} introduces a shared state channel across agent runs, and \texttt{require\_approval="always"} specifies an approval constraint on the hosted email capability. These dependencies are semantically meaningful, but they are not directly represented as traditional function calls or variable-level data flow. As a result, traditional program dependency may capture local control and data dependences inside tool implementations, yet still miss the framework-induced semantics, such as agent-to-tool calling or agent-to-agent handoffs. Hence, analyzing agent programs requires explicitly modeling framework-defined semantics and recovering agent-specific dependencies. 

\noindent\textbf{Challenges.}
However, constructing agent dependencies from source code raises three challenges.
\begin{itemize}[leftmargin=10pt]
    \item \textit{Framework-specific Abstractions.} Different frameworks encode similar agent concepts through different APIs. In the example, an OpenAI Agents-style program represents tools through \texttt{@function\_tool} and \texttt{HostedMCPTool}, agent behavior through \texttt{Agent} constructors, inter-agent transfer through \texttt{handoff}, and tool approval through MCP tool metadata. A LangGraph program may express similar behavior through graph nodes, edges, state objects, and tool nodes, while a CrewAI program may express it through agents, tasks, crews, tools, and delegation. The analysis must therefore not be tied to any single framework's syntax and semantics, but instead abstract framework-specific constructs into a unified intermediate representation.

    \item \textit{Agent-specific Implicit Dependencies.} Many dependencies in agent programs are introduced by framework-defined semantics. In the example, \texttt{tools=[...]} declares capability availability and thereby enables agent-to-tool control transfer, while \texttt{handoff(...)} defines a possible agent-to-agent execution transfer and an associated data channel. The shared \texttt{session} further allows information to persist across agent execution. Static analysis must therefore recover the implicit control-flow and data-flow relations that arise between agents and tools, and between one agent and another, under the framework semantics.

    \item \textit{Uncertain Agent Behaviors.} The concrete behavior of an agent program depends partly on runtime-generated model outputs, so static analysis cannot predict a single deterministic execution trace. In the example, it cannot determine whether \texttt{ResearchAgent} will invoke \texttt{WebSearch}, whether execution will be handed off to \texttt{OpsAgent}, or whether \texttt{OpsAgent} will invoke \texttt{SendEmail}. Instead, static analysis must over-approximate the possible execution behavior permitted by the program dependency.
\end{itemize}

\begin{figure*}[t]
\centering
\includegraphics[width=0.99\linewidth]{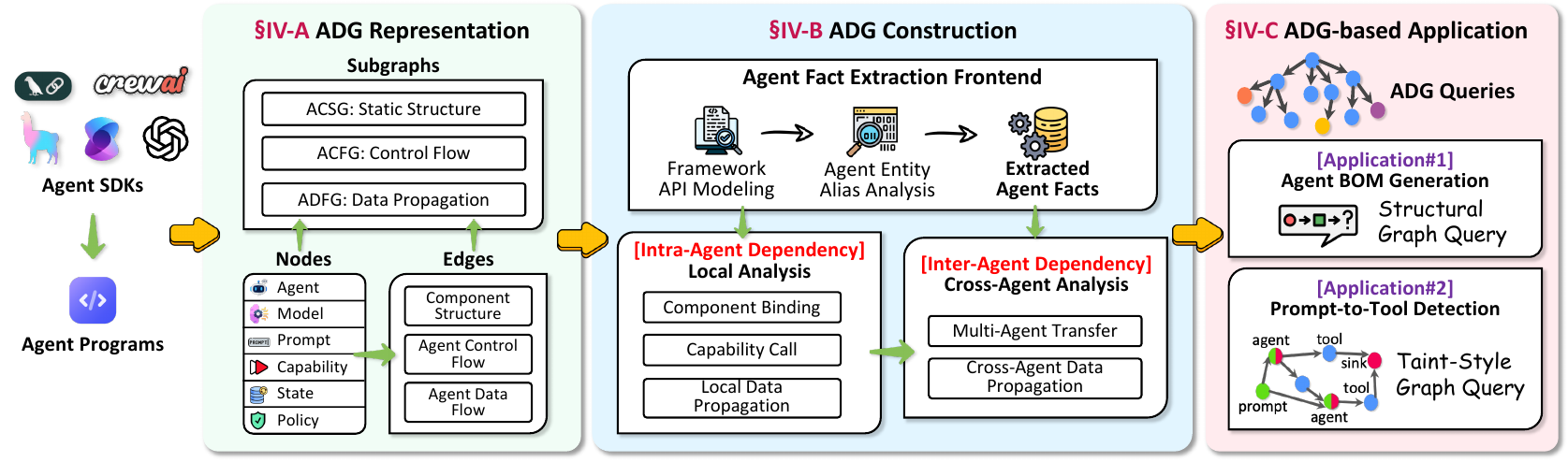}
\caption{Overview of \toolname.}
\label{fig:methodology}
\end{figure*}

\noindent\textbf{Key Insight: Agent Dependency Graph (ADG).}
To address these challenges, \toolname represents an agent program as an Agent Dependency Graph~(ADG), a unified graph abstraction that makes framework-defined agent dependencies explicit. As shown in \autoref{fig:motivating-agent-dependency-example}, ADG normalizes framework-specific constructs into agent entities, including agents, prompts, models, capabilities, states, and policies, and connects them with three kinds of dependency edges. Component-dependency edges capture how agents are assembled from prompts, models, tools, MCP capabilities, states, and policies. Control-flow edges capture possible execution transfers from agents to capabilities and between agents, such as tool calls and handoffs. Data-flow edges capture possible information propagation through prompts, tool arguments and returns, shared sessions, and inter-agent messages. Since the concrete behavior of an agent may depend on runtime model outputs, ADG over-approximates the dependencies permitted by the program structure and framework semantics rather than predicting a single execution trace. In this way, ADG provides a common intermediate representation for analyzing agent programs, and the next section formally defines ADG and describes how \toolname constructs it from source code.
\section{Methodology}
\label{sec:methodology}

This section presents the design of \toolname. As illustrated in \autoref{fig:methodology}, given the source code of an agent program, \toolname statically constructs an Agent Dependency Graph~(ADG), a framework-agnostic graph representation that captures agent-specific dependencies among agents, prompts, models, capabilities, memory, and control policies. Built on ADGs, \toolname further supports various queries and applications, including Agent BOM generation and prompt-to-tool risk analysis.

\subsection{ADG Representation}
\label{sec:methodology-adg}

\noindent\textbf{Syntax of ADGs.}
For an agent program $P$, \toolname represents its dependencies as an ADG,
\[
\mathsf{ADG}_P
=
\langle
\mathsf{ACDG}_P,
\mathsf{ACFG}_P,
\mathsf{ADFG}_P
\rangle .
\]
Here, $\mathsf{ACDG}_P$, $\mathsf{ACFG}_P$, and $\mathsf{ADFG}_P$ denote the Agent Component Dependency Graph~(ACDG), the Agent Control Flow Graph~(ACFG), and the Agent Data Flow Graph~(ADFG). They share a finite node set $V$ and respectively capture static component bindings, control flow transitions, and data flow propagation in $P$.
The node set $V$ is drawn from an abstract agent entity domain $\mathcal{V}$:
\[
\mathcal{V}
=
\mathcal{A}
\uplus
\mathcal{I}
\uplus
\mathcal{M}
\uplus
\mathcal{C}
\uplus
\mathcal{S}
\uplus
\mathcal{G}.
\]
Here, $\mathcal{A}$, $\mathcal{I}$, $\mathcal{M}$, $\mathcal{C}$, $\mathcal{S}$, and $\mathcal{G}$ denote the abstract domains of agent units, prompt contexts, model units, agent capabilities, memory states, and control policies.

\noindent\textbf{Agent Component Dependency Graph.}
The ACDG captures static component binding relations. ACDG is an undirected graph, $\mathsf{ACDG}_P=(V,E_{\mathsf{acdg}})$, where $E_{\mathsf{acdg}}$ contains undirected dependency edges between statically associated entities. Formally,
\[
\begin{aligned}
E_{\mathsf{acdg}}
\subseteq\;&
\big\{
\{a,x\}
\mid
a\in\mathcal{A},
x\in \mathcal{I}\cup\mathcal{M}\cup\mathcal{C}\cup\mathcal{S}
\big\}\\
&\cup
\big\{
\{g,x\}
\mid
g\in\mathcal{G},
x\in \mathcal{A}\cup\mathcal{C}
\big\}\\
&\cup
\big\{
\{a_1,a_2\}
\mid
a_1,a_2\in\mathcal{A}
\big\}.
\end{aligned}
\]
The first class records that an agent is statically bound to its prompt context, model, capabilities, and memory state. The second class records that a control policy is attached to an agent or capability. The third class records orchestration workflow between agent units.

\begin{figure*}[t]
\centering

\begin{minipage}[t]{0.30\textwidth}
\vspace{0pt}
\begin{sourcebox}
session = SQLiteSession(
  ResearchMem, mem.db)

@function_tool
def web_search(query: str) -> str:
  return ...

send_email = HostedMCPTool(
  {server_label: send_email,
   server_url: ...},
  tool_config={
   require_approval: always})

ops_agent = Agent(
  name=OpsAgent,
  model=gpt-4o,
  instructions=...,
  tools=[send_email])

research_agent = Agent(
  name=ResearchAgent,
  model=gpt-4o,
  instructions=...,
  tools=[web_search],
  handoffs=[
    handoff(agent=ops_agent)])

result = await Runner.run(
  research_agent, user_input,
  session=session)
\end{sourcebox}
\end{minipage}
\hfill
\begin{minipage}[t]{0.27\textwidth}
\vspace{0pt}
\begin{resolvebox}
web_search
  -> function_tool
  -> capability: WebSearch

send_email
  -> HostedMCPTool(...)
  -> capability: SendEmail
  -> policy: require_approval

ops_agent
  -> Agent(OpsAgent)
  -> model: gpt-4o
  -> prompt: OpsInst
  -> tools: [SendEmail]

research_agent
  -> Agent(ResearchAgent)
  -> model: gpt-4o
  -> prompt: ResearchInst
  -> tools: [WebSearch]
  -> handoff target: OpsAgent

session
  -> SQLiteSession(ResearchMem)
  -> state: ResearchMem

Runner.run(...)
  -> entry: ResearchAgent
  -> input: UserInput
  -> session: ResearchMem
\end{resolvebox}
\end{minipage}
\hfill
\begin{minipage}[t]{0.40\textwidth}
\vspace{0pt}
\begin{factbox}
\affont

\factsection{Entity Definition Facts}
\begin{factrecords}
\fact{AgentDef} &
  \obj{ResearchAgent} & \obj{research\_agent} \\
\fact{PromptDef} &
  \obj{ResearchInst} & \obj{...} \\
\fact{ToolDef} &
  \obj{WebSearch} & \obj{custom\_tool} \\
\fact{MCPDef} &
  \obj{SendEmail} & \obj{MCP\_URL} \\
\fact{StateDef} &
  \obj{ResearchMem} & \obj{mem.db} \\
\fact{PolicyDef} &
  \obj{ApprovalPolicy} & \obj{require\_approval} \\
\fact{$\cdots$} &
  \obj{$\cdots$} & \obj{$\cdots$} \\
\end{factrecords}

\factsection{Component Structure Facts}
\begin{factrecords}
\fact{BindModel} &
  \obj{ResearchAgent} & \obj{gpt-4o} \\
\fact{BindPrompt} &
  \obj{ResearchAgent} & \obj{ResearchInst} \\
\fact{BindTool} &
  \obj{ResearchAgent} & \obj{WebSearch} \\
\fact{BindMCP} &
  \obj{OpsAgent} & \obj{SendEmail} \\
\fact{BindState} &
  \obj{ResearchAgent} & \obj{ResearchMem} \\
\fact{$\cdots$} &
  \obj{$\cdots$} & \obj{$\cdots$} \\
\end{factrecords}

\factsection{Control Flow Facts}
\begin{factrecords}
\fact{Call} &
  \obj{ResearchAgent} & \obj{WebSearch} \\
\fact{Transfer} &
  \obj{ResearchAgent} & \obj{OpsAgent} \\
\fact{Call} &
  \obj{OpsAgent} & \obj{SendEmail} \\
\fact{BranchIf} &
  \obj{ApprovalPolicy} & \obj{SendEmail} \\
\fact{$\cdots$} &
  \obj{$\cdots$} & \obj{$\cdots$} \\
\end{factrecords}

\factsection{Data Flow Facts}
\begin{factrecords}
\fact{ReadState} &
  \obj{ResearchMem} & \obj{ResearchAgent} \\
\fact{WriteState} &
  \obj{ResearchAgent} & \obj{ResearchMem} \\
\fact{ToolCallArg} &
  \obj{OpsAgent} & \obj{SendEmail} \\
\fact{ToolCallRet} &
  \obj{SendEmail} & \obj{OpsAgent} \\
\fact{InterAgentMsg} &
  \obj{ResearchAgent} & \obj{OpsAgent} \\
\fact{$\cdots$} &
  \obj{$\cdots$} & \obj{$\cdots$} \\
\end{factrecords}
\end{factbox}
\end{minipage}

\vspace{2pt}
\caption{Agent fact extraction. \toolname resolves framework object references in an OpenAI-Agent-SDK style code excerpt and normalizes them into entity, component, control, and data facts for ADG construction.}
\label{fig:fact-extraction-example}
\end{figure*}

\noindent\textbf{Agent Control Flow Graph.}
The ACFG captures agent control-flow transitions. ACFG is a directed graph $\mathsf{ACFG}_P=(V,E_{\mathsf{acfg}})$, where $E_{\mathsf{acfg}}$ is induced from the transition relation
\[
T_{\mathsf{acfg}}
\subseteq
\mathcal{A}
\times
(\mathcal{G}\cup\{\bot\})
\times
(\mathcal{A}\cup\mathcal{C}).
\]
A tuple $(a,\bot,t)\in T_{\mathsf{acfg}}$ means that execution may transfer directly from agent $a$ to target $t$, where $t$ is either another agent or a capability. A tuple $(a,g,t)\in T_{\mathsf{acfg}}$ means that the transfer from $a$ to $t$ is guarded by a control policy object $g$.
The edge set $E_{\mathsf{acfg}}$ is derived from $T_{\mathsf{acfg}}$ as follows:
\[
\begin{aligned}
(a,\bot,t) & \in  T_{\mathsf{acfg}}
\Rightarrow
a\rightarrow t\in E_{\mathsf{acfg}},\\
(a,g,t)\in T_{\mathsf{acfg}}
&\Rightarrow
a\rightarrow g\in E_{\mathsf{acfg}}
\land
g\rightarrow t\in E_{\mathsf{acfg}}.
\end{aligned}
\]

\noindent\textbf{Agent Data Flow Graph.}
The ADFG is a directed graph, $\mathsf{ADFG}_P = (V,E_{\mathsf{adfg}})$.
It records possible information propagation among agent-program entities:
\[
\begin{aligned}
E_{\mathsf{adfg}}
\subseteq\;&
E^{\mathsf{ctx}}_{\mathcal{I}\rightarrow\mathcal{A}}
\cup
E^{\mathsf{arg}}_{\mathcal{A}\rightarrow\mathcal{C}}
\cup
E^{\mathsf{ret}}_{\mathcal{C}\rightarrow\mathcal{A}}\\
&
\cup
E^{\mathsf{msg}}_{\mathcal{A}\rightarrow\mathcal{A}}
\cup
E^{\mathsf{write}}_{\mathcal{A}\rightarrow\mathcal{S}}
\cup
E^{\mathsf{read}}_{\mathcal{S}\rightarrow\mathcal{A}} .
\end{aligned}
\]
Here, $i\rightarrow a$ means that a prompt context may influence an agent. The pair of edge families $a\rightarrow c$ and $c\rightarrow a$ model capability arguments and capability returns. The edge family $a_1\rightarrow a_2$ models message, task-result, or handoff-payload propagation between agents. The pair of edge families $a\rightarrow s$ and $s\rightarrow a$ models memory state writes and reads. 

\subsection{ADG Construction}
\label{sec:methodology-construction}

As shown in \autoref{fig:methodology}, ADG construction consists of three stages. 
First, the framework-specific frontend analyzer extracts normalized agent facts from agent-framework constructs. Second, the intra-agent dependency analyzer derives local component bindings, capability calls, and data propagation inside each agent. Third, the inter-agent dependency analyzer derives multi-agent transfers and cross-agent data propagation.

\noindent\textbf{Agent Fact Extraction Frontend.}
The first stage of \toolname is a fact extraction frontend, which translates framework-specific source code into normalized agent facts. Although agent programs are written in a general-purpose host language, their agent-level semantics are often specified through framework-defined constructs, such as agent constructors, tool decorators, hosted tool objects, handoff declarations, workflow commands, and session objects. Therefore, the frontend first recognizes these constructs and then resolves the alias references needed to recover agent dependencies.

\begin{table}[t]
\centering
\caption{Framework-specific constructs modeled and supported by the fact extraction frontend.}
\label{tab:agent-fact-frontend}
\small
\setlength{\tabcolsep}{5pt}
\renewcommand{\arraystretch}{1.08}
\begin{threeparttable}
\begin{tabular}{lrrrrr}
\toprule
\textbf{Framework} &
\textbf{Entity} &
\textbf{Bind.} &
\textbf{Ctrl.} &
\textbf{Data} &
\textbf{Total} \\
\midrule
OpenAI Agents SDK   & 19 & 17 & 23 & 12 & 48 \\
LangChain/LangGraph & 17 & 14 & 22 & 3  & 27 \\
CrewAI              & 6  & 6  & 8  & 4  & 15 \\
LlamaIndex          & 15 & 9  & 6  & 7  & 22 \\
Semantic Kernel     & 25 & 10 & 2  & 6  & 31 \\
\midrule
\textbf{Total}      & 82 & 56 & 61 & 32 & 143 \\
\bottomrule
\end{tabular}
\begin{tablenotes}[flushleft]
\footnotesize
\item \textit{Note.} One construct may contribute to more than one semantic category. For example, LangGraph's \texttt{Command(update=..., goto=...)} contributes to data flowvia \texttt{update} and control flow via \texttt{goto}.
\end{tablenotes}
\end{threeparttable}
\end{table}

As shown in \autoref{fig:fact-extraction-example}, \toolname analyzes an OpenAI-Agent-SDK style program in three steps. 
First, it recognizes constructs that introduce agent entities and dependencies. For example, \texttt{@function\_tool} registers \texttt{web\_search} as a callable tool, \texttt{HostedMCPTool} creates the MCP capability with an approval policy, and \texttt{Agent(...)} binds models, instructions, and tools to an agent. 
Second, \toolname resolves the alias references to connect each use of an agent construct back to its original definition using existing static analysis tools~\cite{wang2026yasa}. For instance, \texttt{send\_email} is resolved from the tool list of \texttt{OpsAgent} to its \texttt{HostedMCPTool} definition. 
% To resolve such references, \toolname uses a worklist-based procedure that starts from recognized framework constructs and iteratively follows framework-relevant assignments, keyword arguments, containers, object attributes, tool lists, handoff targets, and session objects until no new referenced definitions are discovered. 
Third, the resolved objects are represented as opcode-and-operand-style agent facts, including entity facts, component-dependency facts, control-flow facts, and data-flow facts.
\autoref{tab:agent-fact-frontend} summarizes the framework-specific constructs currently modeled and supported by the \toolname. We group their induced semantics into four categories: entity definition, component binding, control flow, and data flow. In total, \toolname models and supports 143 constructs across five agent frameworks. 

\noindent\textbf{Intra-agent Local Analysis.}
After fact extraction, \toolname performs intra-agent local analysis to initialize dependencies inside each agent. This step consumes facts whose entities are locally associated with the same agent. Entity facts, such as \texttt{AgentDef}, \texttt{ToolDef}, and \texttt{MCPDef}, introduce ADG nodes. Component binding facts, such as \texttt{BindModel}, \texttt{BindPrompt}, and \texttt{BindTool}, introduce ACDG edges between an agent and its bound components.
\toolname then analyzes intra-agent control dependencies. A fact \texttt{Call(a,c)} indicates that agent $a$ may invoke capability $c$, and therefore induces an ACFG edge from $a$ to $c$. If the invocation is protected by a control policy, \toolname interprets the corresponding \texttt{BranchIf} fact as a conditional jump in ACFG. For example, \texttt{Call(OpsAgent, SendEmail)} together with \texttt{BranchIf(ApprovalPolicy, SendEmail)} is initialized as a guarded control path from \texttt{OpsAgent} to \texttt{SendEmail} through \texttt{ApprovalPolicy}.
Finally, \toolname resolves intra-agent data dependencies. Prompt bindings induce ADFG edges from prompt contexts to agents. Tool-call arguments and return facts, such as \texttt{ToolCallArg} and \texttt{ToolCallRet}, induce data-flow edges from agents to capabilities and from capabilities back to agents.

\noindent\textbf{Inter-agent Dependency Analysis.}
The inter-agent dependency analysis constructs dependencies that cross agent boundaries. Such dependencies are introduced by framework constructs for multi-agent orchestration. 
For agent transition control flow, a transfer fact such as \texttt{Transfer(a\_1,a\_2)} induces an ACFG edge from source agent $a_1$ to target agent $a_2$. \toolname supports both explicit workflow edges and dynamic routing constructs that enumerate possible next agents.
For cross-agent data flow, \toolname distinguishes explicit and implicit propagation. Explicit propagation is introduced by messages, task outputs, and handoff payloads. For example, an inter-agent message fact \texttt{InterAgentMsg(a\_1,a\_2)} induces an ADFG edge from $a_1$ to $a_2$, indicating that information produced by the source agent may influence the target agent. Implicit propagation is introduced by shared memory or state objects. If agent $a_1$ writes to a state object $s$ and agent $a_2$ reads from the same object, \toolname represents the cross-agent flow as the ADFG path $a_1 \rightarrow s \rightarrow a_2$. 
Overall, \toolname adopts an over-approximate design. The resulting ADG represents possible dependencies rather than a single concrete execution trace.

\subsection{ADG-based Application}
\label{sec:methodology-application}

\toolname supports downstream analyses by expressing them as graph queries over ADGs. In this paper, we demonstrate two representative applications: (1) Agent BOM generation~\cite{li2026agentbom,dutta2026agentriskbom} and (2) prompt-to-tool risk detection~\cite{shen2025llmvul,fariha2026llmvul,wang2025llmscvul}. The former queries ACDG to traverse the components bound to each agent, while the latter combines ADFG and ACFG to identify whether prompt-influenced data may reach the arguments of a privileged capability.

\begin{figure}[t]
\centering
\fbox{
\begin{minipage}{0.95\linewidth}
\footnotesize
\setlength{\abovedisplayskip}{3pt}
\setlength{\belowdisplayskip}{3pt}

\textbf{Notations.}
\begin{itemize}[leftmargin=12pt,itemsep=1pt,topsep=2pt]
    \item $P$: an agent program.
    \item $V$: the shared node set of $\mathsf{ADG}_P$.
    \item $u\leadsto_{\mathsf{str}}v$: $v$ is reachable from $u$ through one or more ACSG edges.
    \item $u\leadsto_{\mathsf{ctrl}}v$: $v$ is reachable from $u$ through one or more ACFG edges.
    \item $u\leadsto_{\mathsf{data}}v$: $v$ is reachable from $u$ through one or more ADFG edges.
    \item $E_{\mathsf{arg}}$: agent-to-capability data flow edges, i.e., $E^{\mathsf{arg}}_{\mathcal{A}\rightarrow\mathcal{C}}$.
    \item $\mathsf{Src}_{\mathsf{prompt}}\subseteq\mathcal{I}$: prompt or instruction sources.
    \item $\mathsf{Snk}_{\mathsf{tool}}\subseteq\mathcal{C}$: sensitive capability sinks.
\end{itemize}

\textbf{Agent BOM Query.}
\[
\begin{aligned}
\mathsf{BOM}(P)
=
\{(a,x)\mid\;&
a\in \mathcal{A},
x\in\mathcal{I}\cup\mathcal{M}\cup\mathcal{C}\cup\mathcal{S}\cup\mathcal{A},
a\leadsto_{\mathsf{str}}x
\}.
\end{aligned}
\]

\textbf{Prompt-to-Tool Query.}
\[
\begin{aligned}
R_{\mathsf{pa}}
=
\{(p,a)\mid\;
p\in\mathsf{Src}_{\mathsf{prompt}},\;
a\in\mathcal{A},
p\leadsto_{\mathsf{data}}a
\}.
\end{aligned}
\]
\[
\begin{aligned}
R_{\mathsf{ac}}
=
\{(a,c)\mid\;
a\in\mathcal{A},\;
c\in\mathsf{Snk}_{\mathsf{tool}},
a\leadsto_{\mathsf{ctrl}}c
\}.
\end{aligned}
\]
\[
\begin{aligned}
\mathsf{P2T}(P)
=
\{(p,a,c)\mid\;
(p,a)\in R_{\mathsf{pa}},
(a,c)\in R_{\mathsf{ac}},
(a,c)\in E_{\mathsf{arg}}
\}.
\end{aligned}
\]

\end{minipage}
}
\caption{Queries for Agent BOM and Prompt-to-Tool.}
\label{fig:adg-query-rules}
\end{figure}

\noindent\textbf{Agent BOM Generation.}
The first application generates an Agent BOM for an agent program. Given the ADG of a program $P$, \toolname queries ACDG to collect the components structurally associated with each agent. As shown in \autoref{fig:adg-query-rules}, $\mathsf{BOM}(P)$ returns pairs $(a,x)$, where $a$ is an agent node and $x$ is the prompt, model, capability, memory state, or another agent from $a$ through ACDG. 

\noindent\textbf{Prompt-to-Tool Detection.}
The second application detects whether prompt-controlled information may influence a sensitive capability invocation. As shown in \autoref{fig:adg-query-rules}, $\mathsf{P2T}(P)$ first derives $R_{\mathsf{pa}}$, which contains prompt-source-to-agent pairs reachable through ADFG, and $R_{\mathsf{ac}}$, which contains agent-to-sensitive-capability pairs reachable through ACFG. A reported triple $(p,a,c)$ requires $(p,a)\in R_{\mathsf{pa}}$, $(a,c)\in R_{\mathsf{ac}}$, and $(a,c)\in E_{\mathsf{arg}}$. The first condition captures prompt influence on the agent context, the second checks that the sensitive capability is executable under framework control flow, and the third ensures that the influenced agent context may enter the capability arguments. Since $\leadsto_{\mathsf{data}}$ is transitive, the query covers iterative taint propagation through multiple agents.
\section{Evaluation}
\label{sec:evaluation}

In this section, we evaluate \toolname{} with the following research questions~(RQs):

\noindent\hangindent=2.5em\hangafter=1
\textbf{RQ1 [ADG Construction]} 
How effectively does \toolname{} recover agent entities and dependencies from framework-specific constructs, compared with existing agent static analysis tools?

\noindent\hangindent=2.5em\hangafter=1
\textbf{RQ2 [Agent BOM]} 
How effectively does \toolname{} generate Agent BOMs, in terms of agent-specific components and binding relationships?

\noindent\hangindent=2.5em\hangafter=1
\textbf{RQ3 [Prompt-to-Tool Risk]} 
What prompt-to-tool risks can \toolname{} detect in real-world agent programs, and how reliable are the detected reports?

\noindent\hangindent=2.5em\hangafter=1
\textbf{RQ4 [Scalability and Performance]} 
How efficiently does \toolname{} analyze real-world agent programs?

\subsection{Experimental Setup}
\label{sec:evaluation-setup}

\noindent\textbf{Implementation.}
We have implemented \toolname{} using 11.4 KLoC of Python, 4.5 KLoC of TypeScript checker code, and 2.9 KLoC of framework semantic registry specifications. The fact extraction frontend is built on YASA~\cite{wang2026yasa}, reusing its resolved callees, call arguments, symbolic values, and source locations to emit normalized agent facts through framework-specific semantic rules. The current registry covers five frameworks, including OpenAI Agents SDK, LangChain/LangGraph, CrewAI, LlamaIndex, and Semantic Kernel, and models 143 framework-specific constructs.

\noindent\textbf{Running Environment.}
All experiments were conducted on a server running Ubuntu Linux 24.04.3 LTS. The server is equipped with two AMD EPYC 9554, 128 physical CPU cores, 1007~GiB RAM, five NVIDIA A100 80GB GPUs, eight 7~TB SSDs, and four 14.6~TB HDDs. 

\noindent\textbf{Baselines.}
We select baselines according to the evaluation task. For RQ1, we compare \toolname{} with two agent static analysis tools, \textsc{Agent-Wiz}~\cite{agentwiz} and \textsc{AgenticRadar}~\cite{agentic-radar}, which are the closest tools for recovering agent entities and framework-level dependencies from agent programs. For RQ2, we compare ADG-based Agent BOM generation with three existing AI/Agent BOM systems, including Trusera \textsc{Trusera-AI-BOM}~\cite{trusera_aibom}, Drako \textsc{Drako-Agent-BOM}~\cite{drako_agent_bom}, and Cisco \textsc{Cisco-AI-BOM}~\cite{cisco_aibom}. Since they differ in output schema and BOM scope, we compare them qualitatively by component coverage and quantitatively using statistics of generated components and relationships. 
% For RQ3, we compare prompt-to-tool risk detection with two static agent-security baselines, \textsc{agentic-guard}~\cite{agentic-guard} and \textsc{Agent Audit}~\cite{zhang2026agentaudit}, which provide rule-based security analysis for agent programs. We run all runnable baselines with their default configurations. We also consider \textsc{TaintP2X}~\cite{icse26taintp2x}, \textsc{LLMSmith}~\cite{liu2024llm4shell}, and \textsc{AgentFuzz}~\cite{liu2025agentfuzz}, but do not include them as direct baselines. \textsc{TaintP2X} targets LLM-integrated applications rather than framework-defined agent programs, while \textsc{LLMSmith} and \textsc{AgentFuzz} are fuzz testing rather than static analyzers.

\begin{table*}[t]
\centering
\caption{Qualitative comparison of agent dependency modeling capabilities.}
\label{tab:agent-qualitative}
\resizebox{0.9\textwidth}{!}{%
\begin{tabular}{ccccccccc}
\hline
\textbf{Tool} &
\textbf{Support} &
\textbf{IR \& Analysis} &
\textbf{Capability} &
\textbf{Node} &
\textbf{Control} &
\textbf{Data} &
\textbf{Memory} & 
\textbf{Policy} \\
\hline

\textsc{Agentic Radar}
& 5 frameworks
& AST-based Parsing
& Tool, MCP
& 5 types
& \half
& \none
& \none
& \half \\

\textsc{Agent-Wiz}
& 10 frameworks
& AST-based Parsing
& Tool
& 4 types
& \half
& \none
& \half
& \none \\

\hline
\toolname{}
& 5 frameworks
& ADG \& Alias Resolution
& Tool, MCP, Skills
& 7 types
& \full
& \full
& \half
& \full \\

\hline
\end{tabular}}
\end{table*}

\begin{table*}[t]
\centering
\caption{Analysis outcomes and recovered agent dependency nodes and edges on \textit{ADG-Eval}.}
\label{tab:rq1-quantitative}
\resizebox{\textwidth}{!}{%
\begin{threeparttable}
\begin{tabular}{l|rrr|rrrrrr|rrr|rr}
\hline
\textbf{Tool} &
\textbf{Success} &
\textbf{TO} &
\textbf{Fail} &
\textbf{Agent} &
\textbf{Cap.} &
\textbf{Prompt} &
\textbf{Model} &
\textbf{Memory} &
\textbf{Policy} &
\textbf{Struct.} &
\textbf{Ctrl.} &
\textbf{Data} &
\textbf{Med.N} &
\textbf{Med.E} \\
\hline
\textsc{Agent-Wiz}
& 60 & -- & --
& 284 & 35 & -- & 54 & -- & --
& 97 & 185 & --
& 4 & -- \\

\textsc{Agentic Radar}
& 31 & -- & 29
& 149 & 54 & 32 & 41 & -- & --
& 73 & 352 & --
& 2.5 & 2 \\

\hline
\toolname{}
& 59 & 1 & --
& 738 & 314 & 599 & 180 & 39 & 60
& 973 & 639 & 1,222
& 17 & 28 \\
\hline
\end{tabular}

\begin{tablenotes}[flushleft]
\footnotesize
\item \textbf{TO} denotes timeout. \textbf{Cap.} denotes capabilities. \textbf{Struct.}, \textbf{Ctrl.}, and \textbf{Data} denote structure, control, and data dependencies. \textbf{Med.N} and \textbf{Med.E} denote the per-project median numbers of recovered nodes and edges.
\end{tablenotes}
\end{threeparttable}}
\end{table*}

\noindent\textbf{Datasets and Evaluation Methods.}
We use a large-scale corpus of real-world agent programs, named \textit{AgentZoo}, as the basis of our evaluation. We collect Python repositories from GitHub by searching for framework-specific imports of the five supported agent frameworks, including OpenAI Agents SDK, LangChain/LangGraph, CrewAI, LlamaIndex, and Semantic Kernel. We remove forked and archived repositories, filter out repositories without recognized agent-framework constructs, and analyze the latest default-branch snapshot of each remaining repository. This process yields 5,399 projects, including 3,823 LangChain/LangGraph projects, 947 CrewAI projects, 442 OpenAI Agents SDK projects, 146 LlamaIndex projects, and 41 Semantic Kernel projects.
For RQ1, we evaluate agent dependency recovery on a sampled dataset, named \textit{ADG-Eval}. Since \toolname{} and the two agent-analysis baselines support different framework scopes, we focus on the three frameworks commonly supported by all three tools: OpenAI Agents SDK, LangChain/LangGraph, and CrewAI. To avoid biasing the sample toward either small examples or large repositories, we perform stratified sampling by project size within each framework. This process randomly samples 20 projects per framework and 60 projects in total. We run \toolname{}, \textsc{Agent-Wiz}, and \textsc{Agentic Radar} on these projects and compare their outputs in two ways. First, we qualitatively compare their supported analysis capabilities, including IR design, capability modeling, memory/state modeling, policy modeling, and data-flow support. Second, we normalize their outputs into common high-level node and edge categories and report quantitative statistics, including analysis outcomes, recovered node categories, recovered edge categories, and generated node/edge entries. For \toolname{}, two authors further inspect the source code and generated ADGs to identify false positives and potential false negatives.
For RQ2, we evaluate Agent BOM generation on a broader five-framework sample, named \textit{BOM-Eval}. In addition to the 60 projects used in RQ1, we sample 20 LlamaIndex projects and 20 Semantic Kernel projects using the same size-stratified sampling strategy, resulting in 100 projects in total. We run \toolname{} and all available BOM baselines on these projects and compare their generated BOMs qualitatively in terms of asset coverage and quantitatively using statistics of generated components and relationships. For the BOMs generated by \toolname{}, two authors further inspect the source code to identify false positives and potential false negatives under our Agent BOM schema.
For RQ3, we run \toolname{} and all runnable baselines on the full \textit{AgentZoo} and construct a sampled manual audit set, \textit{P2T-Audit}, by labeling 100 reports from \toolname{} and 100 non-reported agents with privileged capabilities to estimate precision and potential false negatives.
For RQ4, we use \textit{AgentZoo} to evaluate scalability and performance, including success rate, timeout/failure rate, end-to-end analysis time, and graph sizes.

\subsection{RQ1: ADG Construction}

\noindent\textbf{Qualitative Comparison.}
\autoref{tab:agent-qualitative} compares the analysis capability of \toolname{} with two representative open-source agent analysis tools. Although \textsc{Agentic Radar} and \textsc{Agent-Wiz} support major framework parsers, their analyses are primarily AST-based and focus on recovering explicit workflow structures. They provide partial support for inter-agent control flow, but do not model the agent-specific data flow dependencies. At the capability level, \textsc{Agentic Radar} supports MCP tools while \textsc{Agent-Wiz} mainly focuses on traditional tool calls. In contrast, \toolname{} performs alias resolution and constructs ADGs, providing deeper analysis on implicit agent dependencies.

\begin{table*}[t]
\centering
\caption{Qualitative comparison of Agent BOM generation approaches.}
\label{tab:rq2-bom-qualitative}
\resizebox{0.8\textwidth}{!}{%
\begin{threeparttable}
\begin{tabular}{l|l|ccccc|ccccc}
\hline
\textbf{\multirow{2}{*}{Tool}} &
\textbf{\multirow{2}{*}{Technique}} &
\multicolumn{5}{c|}{\textbf{Component Coverage}} &
\multicolumn{5}{c}{\textbf{Binding Relationship Coverage}} \\
\cline{3-12}
&
&
\textbf{Agent} &
\textbf{Model} &
\textbf{Prompt} &
\textbf{Cap.} &
\textbf{Memory} &
\textbf{A-C} &
\textbf{A-P} &
\textbf{A-M} &
\textbf{A-Mem.} &
\textbf{A-A} \\
\hline

\textsc{Trusera-AI-BOM}
& \makecell[l]{AST-based Parsing}
& \none & \full & \none & \none & \none
& \none & \none & \none & \none & \none \\

\textsc{Drako-Agent-BOM}
& AST-based Parsing
& \half & \full & \half & \half & \none
& \none & \none & \none & \none & \none \\

\textsc{Cisco-AI-BOM}
& \makecell[l]{AST-based + LLM}
& \half & \half & \none & \none & \none
& \none & \none & \half & \none & \none \\

\hline
\toolname{}
& ADG-based
& \full & \full & \full & \full & \half
& \full & \half & \full & \half & \full \\
\hline
\end{tabular}

\begin{tablenotes}[flushleft]
\footnotesize
\item \textbf{Cap.} denotes callable capabilities, mainly including tools, MCP servers, and skills.
\textbf{A-C}, \textbf{A-P}, \textbf{A-M}, \textbf{A-Mem.}, and \textbf{A-A} denote agent-capability, agent-prompt, agent-model, agent-memory, and agent-agent binding relationships, respectively.
\end{tablenotes}
\end{threeparttable}}
\end{table*}

\begin{table*}[t]
\centering
\caption{Quantitative comparison of generated Agent BOMs on \textit{BOM-Eval}.}
\label{tab:rq2-bom-quantitative}
\resizebox{\textwidth}{!}{%
\begin{tabular}{l|c|rrrrrr|rrrrr|rr}
\hline
\textbf{\multirow{2}{*}{Tool}} &
\textbf{\multirow{2}{*}{Success}} &
\multicolumn{6}{c|}{\textbf{Assets}} &
\multicolumn{5}{c|}{\textbf{Binding Relationships}} &
\textbf{\multirow{2}{*}{Assets}} &
\textbf{\multirow{2}{*}{Rels.}} \\
\cline{3-13}
&
&
\textbf{Agent} &
\textbf{Model} &
\textbf{Prompt} &
\textbf{Cap.} &
\textbf{Mem.} &
\textbf{Other} &
\textbf{A-C} &
\textbf{A-P} &
\textbf{A-M} &
\textbf{A-Mem.} &
\textbf{A-A} &
&
\\
\hline

\textsc{Drako-Agent-BOM}
& 100/100
& 141 & 201 & 275 & 194 & -- & --
& -- & -- & -- & -- & --
& 971 & 107 \\

\textsc{Trusera-AI-BOM}
& 100/100
& -- & 236 & -- & -- & -- & 344
& -- & -- & -- & -- & --
& 3,615 & -- \\

\textsc{Cisco-AI-BOM}
& 100/100
& 17 & 90 & -- & 4 & 3 & 27
& -- & -- & 20 & -- & --
& 141 & 22 \\

\hline

\toolname{}
& 98/100
& 784 & 260 & 434 & 565 & 37 & 195
& 403 & 183 & 254 & 36 & 50
& 2,295 & 1,008 \\

\hline
\end{tabular}
}
\end{table*}

\noindent\textbf{Quantitative Comparison.}
We further evaluate agent dependency recovery on \textit{ADG-Eval}.
As shown in \autoref{tab:rq1-quantitative}, \toolname{} successfully analyzes 59 of 60 projects, with one timeout on a large project. \textsc{Agent-Wiz} successfully runs on all projects but mainly recovers explicit agents, conventional tools, models, and shallow workflow relations. \textsc{Agentic Radar} recovers more control dependencies than \textsc{Agent-Wiz}, mainly due to workflow and handoff extraction, but fails on 29 projects and does not recover memory state, control policy, or data dependencies. At the project level, \toolname{} recovers a median of 17 nodes and 28 edges per project, compared with 4 nodes and no edges for \textsc{Agent-Wiz}, and 2.5 nodes and 2 edges for \textsc{Agentic Radar}. These results show that existing AST-based tools primarily recover direct and explict workflow structures, while ADG-based analysis captures richer agent-specific dependencies.

\subsection{RQ2: Agent BOM}

\noindent\textbf{Qualitative Comparison.}
We compare \toolname{} with existing AI/Agent BOM tools in terms of generation technique, component coverage, and binding-relationship coverage. As shown in \autoref{tab:rq2-bom-qualitative}, existing tools mainly provide AI inventory rather than dependency-aware Agent BOMs. \textsc{Trusera-AI-BOM} primarily reports models, packages, and other AI-related inventory entries, but does not recover agent components or explicit agent bindings. \textsc{Drako-Agent-BOM} recovers some agent-oriented components through AST-based parsing, but its generated BOM also does not provide explicit binding relationships. Without such bindings, developers can see that components exist, but cannot determine how they are connected. \textsc{Cisco-AI-BOM} combines AST parsers with LLM-assisted analysis, but its generated BOM mainly covers model- or embedding-related inventory and only partial agent-model bindings. In contrast, \toolname{} generates Agent BOMs from ADGs, which explicitly record binding dependencies among agent-specific components in the ACDG edges.

\noindent\textbf{Quantitative Comparison.}
We further compare the generated BOM on \textit{BOM-Eval}.
As shown in \autoref{tab:rq2-bom-quantitative}, \toolname{} successfully analyzes 98 of 100 projects and generates 2,295 components and 1,008 binding relationships. 
In contrast, existing BOM tools produce substantially fewer binding relationships. \textsc{Drako-Agent-BOM} successfully generates 971 components, including agents, models, prompts, and capabilities, but it produces no explicit binding relationships. \textsc{Trusera-AI-BOM} generates 3,615 inventory entries, but most of them are model, software package, or other related entries, and it produces no agents, capabilities, prompts, memory state objects, or binding relationships. \textsc{Cisco-AI-BOM} generates 141 components and 22 relationships, but they are concentrated on agent-model bindings. These results show that \toolname{} provides substantially richer Agent BOMs by recovering both agent-specific components and their binding relationships.

\subsection{RQ3: Prompt-to-Tool Risk}
\label{sec:rq3}

\noindent\textbf{Detection Results.}
We evaluate whether \toolname{} can identify prompt-to-tool risks in real-world agent programs. Among 5,399 analyzed projects, \toolname{} identifies 238 projects with P2T risks, accounting for 4.4\% of the analyzed projects. These projects contain 4,357 P2T findings in total, indicating that risky projects often contain multiple prompt-to-tool paths. \autoref{tab:rq3-p2t-summary} shows the framework-level distribution. Most P2T projects come from LangGraph, LangChain AgentExecutor, and OpenAI Agents SDK. We further classify P2T projects by the side effects of the reached sinks. The most common side effect is external sending, followed by file access, administrative mutation, and command/code execution. These categories are non-exclusive, since one project may contain multiple side-effect types. Overall, 196 of the 238 P2T projects contain a single side-effect type, while 42 projects contain multiple side-effect types. Common overlaps include SQL injection plus file access~(11 projects), external send plus file access~(10 projects), execution plus file access~(5 projects), and execution plus external send~(5 projects).

\begin{table}[t]
\centering
\caption{Summary of prompt-to-tool risk on \textit{AgentZoo}.}
\label{tab:rq3-p2t-summary}
\resizebox{0.8\linewidth}{!}{%
\begin{tabular}{llr}
\hline
\textbf{Group} &
\textbf{Metric} &
\toolname{} \\
\hline

\multirow{2}{*}{Overall}
& P2T Projects & 238 \\
& P2T Findings & 4,357 \\
\hline

\multirow{5}{*}{Framework}
& LangChain/LangGraph & 133 \\
& OpenAI Agents SDK & 63 \\
& LlamaIndex & 27 \\
& CrewAI & 10 \\
& Semantic Kernel & 5 \\
\hline

\multirow{4}{*}{Side Effect}
& External Send & 129 \\
& File Access & 74 \\
& SQL Query & 42 \\
& CMD/Code Execute & 41 \\
\hline
\end{tabular}}
\end{table}

\noindent\textbf{Manual Validation.}
To assess reliability, we construct \textit{P2T-Audit}, a sampled audit set containing 100 reports from \toolname{} and 100 non-reported agents with privileged capabilities. Two authors manually inspected the source code. Among the 100 sampled reports, 73 of them exist prompt-to-tool vulnerabilities, yielding a precision of 73.0\%. However, 25 of the 27 FPs still contain valid taint propagation paths, but are classified as FPs by the incorrect sink semantics rather than incorrect ADG dependencies. For example, 13 cases involve local file reads or path traversal without write/delete effects, 7 involve read-only SQL operations, and 5 involve restricted calculator-style tools rather than general code execution. These cases suggest that more precise sink semantics, such as distinguishing the read/write modes of \texttt{open}, can further improve the reported precision.
To characterize potential false negatives, we further inspect 100 non-reported agents with privileged capabilities and identify 9 missed risks. These misses are mainly caused by custom command wrappers, dynamic tool binding or custom orchestration, LangGraph-specific state/tool bindings, and custom multi-tool wrappers. We discuss these limitations in \autoref{sec:discussion}.

\subsection{RQ4: Scalability and Performance}
\label{sec:rq4}
We evaluate the scalability of \toolname{} on \textit{AgentZoo}. The median end-to-end analysis time is 14.17 seconds, and the 95th percentile time is 163.54 seconds. Considering that \toolname{} performs alias-analysis-based dependency recovery and graph construction, these costs are acceptable for large-scale auditing and batch analysis.
We further examine the size of the generated analysis artifacts. The median ADG contains 208 nodes and 206 edges, while the 95th percentile contains 721 nodes and 832 edges. The largest generated ADG contains 3,370 nodes and 17,678 edges. The median number of extracted agent facts is 43, with a 95th percentile of 313. These results show that ADGs remain compact for most real-world projects because they capture framework-level agent dependencies rather than expanding all host-language statements. Overall, \toolname{} scales to thousands of real-world agent programs while keeping both analysis time and graph size manageable.
\section{Discussion}
\label{sec:discussion}

\noindent\textbf{Broader Implications.}
ADG provides a foundation for reusable static analysis of agent software. First, ADG exposes agent-specific dependencies that are often hidden behind heterogeneous framework APIs, which makes it possible to build framework-agnostic analyses on top of a common intermediate representation. Second, ADG can support a broader family of security and software-engineering tasks beyond the two applications demonstrated, such as auditing unbounded or misconfigured agent iterations through control-flow cycles, detecting multi-agent collusion risks through cross-agent control and data dependencies, and checking unsafe tool-composition patterns through agent-capability queries. Third, ADG can be integrated into hybrid analysis for agent programs. ADG-based static analysis can first identify feasible agent-capability paths and high-risk components, and dynamic techniques such as directed fuzzing can then exercise these paths to validate exploitability and reduce false positives. More broadly, we hope ADG encourages the community to treat agent systems as analyzable software artifacts, enabling systematic and extensible analyses for emerging agent ecosystems.

\noindent\textbf{Limitations.}
\toolname has several limitations. First, as a static analysis framework, it over-approximates agent control and data dependencies and captures dependencies permitted by program structure and framework semantics rather than concrete executions, which may introduce false positives. Second, \toolname currently focuses on framework-defined semantics and is less effective for user-defined agent structures that do not follow recognizable framework APIs or fixed syntactic conventions, since their semantics are often project-specific. Third, the current implementation targets Python-based programs and five agent frameworks. Supporting more languages or frameworks requires additional frontend extractors. Fourth, agent frameworks evolve rapidly, so \toolname's semantic registry should be updated as the framework syntax changes. Finally, ADG captures high-level framework semantics and does not replace low-level program dependency analysis. For tasks that require reasoning inside tool implementations or library calls, ADG can be combined with points-to analysis to track the full taint propagation paths.
\section{Related Work}
\label{sec:related-work}

\noindent\textbf{Static Analysis for Different Frameworks.}
Modern software frameworks often express program semantics through framework-specific constructs, which makes general-purpose static analysis insufficient. Prior work has addressed this challenge across different software ecosystems. Android analyzers such as \textsc{FlowDroid}~\cite{steven2014flowdroid} and \textsc{IccTA}~\cite{li2015iccta} model lifecycle callbacks, inter-component communication, and framework-dispatched entry points for call-graph and information-flow analysis. Spring analyses, such as \textsc{Jasmine}~\cite{chen2023jasmine}, \textsc{Tai-e}~\cite{tan2023taie}, and \textsc{YASA}~\cite{wang2026yasa} model dependency injection, annotations, reflection, and aspect-oriented programming to recover framework-induced call and data flows. Some works also focus on GitHub CI/CD workflow~\cite{siddharth2023argus,wang2026taintawi} and model workflow specifications, action steps, contexts, and script bridges for security analysis. Similar framework-induced feature modeling has also been applied to front-end and mobile frameworks, including React application analysis~\cite{guo2024reactappscan} and MiniApp analysis~\cite{wang2023taintmini,wang2025miniscope,meng2023wemint}. \toolname follows this line of static analysis of framework semantics, but targets agent programs and models agent-specific dependencies.

\noindent\textbf{Program Analysis for LLM Agents.}
Recent work has explored program analysis for LLM-integrated applications and agent programs. Taint-style systems such as \textsc{TaintP2X}~\cite{icse26taintp2x} analyze prompt-to-sink dependencies in LLM-integrated applications, where LLM calls are embedded into conventional application code. However, they do not directly analyze framework-defined agent programs. Other security analyses, such as \textsc{LLMSmith}~\cite{liu2024llm4shell} and \textsc{AgentFuzz}~\cite{liu2025agentfuzz}, use greybox fuzzing to expose vulnerabilities in LLM applications. However, their static components often rely on general-purpose engines such as \textsc{CodeQL}. Existing agent-oriented static tools, such as \textsc{AgenticRadar}~\cite{agentic-radar} and \textsc{AgentWiz}~\cite{agentwiz}, primarily rely on AST-level parsing to recover agents, tools, and workflow configurations. In contrast, \toolname advances this area by introducing ADG as an intermediate representation that explicitly models agent-specific dependencies.
\section{Conclusion}
\label{sec:conclusion}

This paper presents \toolname, the first static analysis framework for recovering and analyzing agent dependencies in agent programs. Specifically, \toolname constructs ADGs, which provides a framework-agnostic graph representation of agents, prompts, models, capabilities, memory states, control policies, and their component-structure, control-flow, and data-flow dependencies. Built on ADGs, \toolname supports dependency-aware Agent BOM generation and taint-style prompt-to-tool risk detection. Our evaluation on 5,399 real-world agent programs shows that \toolname recovers richer agent dependencies than \textsc{Agentic Radar} and \textsc{Agent-Wiz}, generates more informative Agent BOMs than \textsc{Drako-Agent-BOM}, \textsc{Trusera-AI-BOM}, and \textsc{CISCO-AI-BOM}, and identifies 238 real-world prompt-to-tool risks. These results demonstrate that \toolname provides a practical foundation for understanding, governing, and securing emerging agent programs and agent ecosystems.

\newpage
\balance
\bibliographystyle{IEEEtran}
\bibliography{main}

\end{document}